\documentstyle[12pt]{article}

\author{V.A.Tsokur and Yu.M.Zinoviev
       \thanks{E-mail address: ZINOVIEV@MX.IHEP.SU} \\
        {\it Institute for High Energy Physics} \\
        {\it Protvino, Moscow Region, 142284, Russia}}
\title{$N=2$ Supergravity Models \\ with Gauge Kac-Moody Groups}
\date{April 1996}

\begin{document}

\maketitle

\begin{abstract}
 In this paper we consider a class of models for vector and
hypermultiplets, interacting with $N=2$ supergravity, with gauge
groups being an infinite-dimensional Kac-Moody groups. It is shown
that specific properties of Kac-Moody groups, allowing the
introduction of the vector fields masses without the usual Higgs
mechanism, make it possible to break simultaneously both the
supersymmetry and the gauge symmetry. Also, a kind of inverse Higgs
mechanism can be realized, that is, in the considered model there
exists a possibility to lower masses of the scalar fields, which
usually acquire huge masses as a result of supersymmetry breaking.
That allows one to use them, for example, as Higgs fields at the
second step of the gauge symmetry breaking in the unified models. 
\end{abstract}

\newpage

\section*{Introduction}

One of the most serious problems, which arises when one deals with
the exploration of the phenomenological supergravity models, is the
problem of the simultaneous breaking of the supersymmetry and the
gauge symmetry. Attempts to break the gauge symmetry by means of the
usual Higgs mechanism often fail, because both in $N=1$ supergravity
models and in extended supergravity ones all the particles, which
could play the role of the Higgs particles, acquire as a rule masses
of the order of supersymmetry breaking scale. And if in the case of
$N=1$ supergravity in some models it turns out to be possible to
obtain spontaneous gauge symmetry breaking due to radiative
corrections, for the extended supergravities, where mass scales of
the supersymmetry breaking and, correspondingly, masses of the Higgs
particles are essentially larger, it hardly works. 

Let us reconsider the possibilities to have spontaneous gauge
symmetry breaking. As is well known, the key element of all models is
the gauge invariant description of massive vector particles, which is
possible due to the introduction of the Goldstone scalar field
with inhomogeneous transformation law. For the Abelian vector field
the Lagrangian has a very simple form
\begin{equation}
{\cal L} = - \frac14 (\partial_\mu A_\nu - \partial_\nu A_\mu)^2 +
\frac{m^2}2 A_\mu^2 - m A_\mu \partial_\mu \phi + \frac12
(\partial_\mu \phi)^2    \label{abel}
\end{equation}
being invariant under the following gauge transformations: $\delta
A_\mu = \partial_\mu \varepsilon$ and $\delta \phi = m \varepsilon$.

If one starts from the analogous Lagrangian and gauge transformations
in the simplest case of the non-abelian $SU(2)$ gauge group:
\begin{eqnarray}
{\cal L}_0 &=& - \frac14 (F_{\mu\nu}^a)^2 + \frac{m^2}2 (A_\mu^a)^2 -
m A_\mu^a \partial_\mu \phi^a + \frac12 (\partial_\mu\phi^a)^2
\nonumber \\
F_{\mu\nu}^a &=& \partial_\mu A_\nu - g \varepsilon^{abc} A_\mu^b
A_\nu^c - (\mu \leftrightarrow \nu)          \nonumber   \\
\delta_0 A_\mu^a &=& (\partial_\mu \delta^{ab} - g \varepsilon^{abc}
A_\mu^b) \varepsilon^c  \qquad  \delta_0 \phi^a = m \varepsilon^a
\end{eqnarray}
and try to complete both the interaction Lagrangian and the
transformation law of the field $\phi$, requiring the full Lagrangian
to be gauge invariant, one will see that there exist two possible
scenarios \cite{Zin83}. If we proceed without introducing any other
scalar fields we necessarily will come to the gauge invariant
description of massive vector fields where scalars realize a
non-linear $\sigma$-model \cite{Fad71}.

 There is another possibility, leading to the ordinary model of the
spontaneous breaking of $SU(2)$ gauge group through the Higgs
mechanism. To obtain the corresponding formulas, one has to introduce
additional scalar field $\chi$ with the transformation law
$\delta\chi=\frac{g}2\phi^a\varepsilon^a$, which together with
the fields $\phi^a$ forms complex $SU(2)$-dublet, thus avoiding a
non-linear realization.

Therefore, apart from the usual Higgs mechanism, one can exploit the
fact, that in the supergravity theories the scalar fields often
realize non-linear $\sigma$-models of the form $G/H$ and the gauging
of the isometries in such models necessarily leads to the gauge
symmetry breaking. Indeed there are examples of the supergravity
models of such a kind (see e.g. \cite{Wit84b,Zin85a} for $N=2$ case),
but in many $N=1$ and in all extended supergravities one deals with
the non-compact groups $G$, moreover the choice of possible gauge
groups is highly restricted.

So, the generalization to the non-abelian case leads either to the
non-linear models, or to the Higgs mechanism and both of these
schemes fail in the extended supergravity models. But really there
exists a third possibility connected with the infinite-dimensional
groups of the Kac-Moody type. Such groups arise in a natural way when
one deals with the compactifications from higher dimensions and also
in attempts to obtain an effective field theory for superstrings
(e.g. \cite{Giv90,Giv91}). But in this paper we will not rely on any
geometric interpretation and will just investigate $N=2$ supergravity
models with the gauge Kac-Moody groups in the same spirit as in the
\cite{Cad88,Cad89,Cho92}. In the next Section we first of all
reproduce the rather well known formulas for the gauge theory based
on the usual affine Kac-Moody groups and consider the introduction of
the mass terms for the appropriate vector fields. All the formulas,
of course, are similar to those we will get if we consider the
five-dimensional Yung-Mills theory and then compactify the fifth
dimension on the circle. But we stress that in sharp contrast with
the finite dimensional gauge groups the introduction of the mass
terms appears to be as simple matter as in the abelian case --- there
is no need in the Higgs fields with any non-trivial potential. This
allows us to construct a generalization of the simplest models we
started with which could mimic the spontaneous gauge symmetry
breaking $G \rightarrow H$, where for example one can have $G=SU(5)$
and $H=SU(3)\otimes SU(2)\otimes U(1)$.

In Section 2, as a preliminary step to the local $N=2$
supersymmetry, we consider the case of the global one. In this, we
choose to work with massive vector multiplets without central
charges. The reason is that in the $N=2$ supergravity the central
charges are necessarily gauged (see, e.g., \cite{Zac78}), the gauge
fields being graviphotons. But the graviphotons play a very essential
role in the spontaneous supersymmetry breaking, so it would be hard
to have simultaneous breaking of the gauge symmetry and
supersymmetry.

In Section 3 we consider an interaction of our globally $N=2$
supersymmetric models with gauge Kac-Moody groups and the $N=2$
supergravity and investigate the possibilities of spontaneous
symmetry breaking in such models. The main results of our
investigations are twofold. First, we show that it is indeed
possible to have simultaneous breaking of gauge as well as
supersymmetries and calculate the mass spectrum that appears after
such a breaking have taken place. Second, we will see that a kind of
inverse Higgs effect arises --- not only the fields which were
massless could gain masses as a result of supersymmetry breaking,
but some of the initially massive fields could become light or even
massless. It is interesting to note that for such a mechanism
to be operative the scale of the gauge symmetry breaking and the one
for the supersymmetry breaking have to be close to each other.

\section{Kac-Moody groups and gauge symmetry breaking}

The affine Kac-Moody algebra without the central charge has the
following commutation relations:
\begin{equation}
[T^a_m, T^b_n] = f^{abc} T^c_{m+n},            \label{1}
\end{equation}
where $n, m \in {\bf Z}$, $T^a_0\in G$ for any semisimple Lie algebra
$G$ with structural constants $f^{abc}$, so $1<a,b,c<{\rm dim}G$. Let
us assume the generators of this algebra to be antihermitian:
\begin{equation}
(T^a_m)^+ = - T^a_{-m}.
\end{equation}

Let us consider a gauge field that lies in the algebra (\ref{1}):
\begin{equation}
{\cal A}_{\mu} = A_{\mu}{}^a_m T^a_{-m}      \qquad 
{\cal A}_{\mu}^+ = -{\cal A}_{\mu}            \qquad
({\cal A}_{\mu}{}^a_m)^* = A_{\mu}{}^a_{-m} 
\end{equation}
The associated field strength has the usual form:
\begin{equation}
{\cal F}_{\mu\nu} = [\nabla_\mu, \nabla_\nu],
\end{equation}
where $\nabla_\mu = \partial_\mu + {\cal A}_\mu$.

Under infinitesimal gauge transformations with parameter
$\varepsilon$ also lying in algebra (\ref{1}) the gauge field
${\cal A}_\mu$ and the field strength ${\cal F}_{\mu\nu}$ transform
as the following:
\begin{eqnarray}
\delta{\cal A}_\mu &=& [\nabla_\mu, \varepsilon] = \partial_\mu
\varepsilon + [{\cal A}_\mu,\varepsilon]          \nonumber    \\
\delta{\cal F}_{\mu\nu} &=& [{\cal F}_{\mu\nu}, \varepsilon].
\end{eqnarray}

The Lagrangian, invariant under these transformations, has the form,
that coincides with the case of the finite dimensional gauge group:
\begin{equation}
{\cal L} = \frac18 {\rm Sp} \{ {\cal F}_{\mu\nu} {\cal F}_{\mu\nu}\},
\end{equation}
where ${\rm Sp} \{T^a_m T^b_n\} = - 2 \delta^{ab} \delta(m+n)$ with
the notation $\delta(m) = \left\{ 0 \quad {\rm at} \quad m\ne 0
\atop 1 \quad {\rm at} \quad m=0\right.$.

Now one can rewrite all the formulas, obtained above, in the
components:
\begin{eqnarray}
{\cal F}_{\mu\nu} &=& F_{\mu\nu}{}^a_m T^a_{-m}   \qquad    \qquad
\varepsilon = \varepsilon^a_m T^a_{-m}                \nonumber  \\
F_{\mu\nu}{}^a_m &=& \partial_\mu A_\nu{}^a_m - \partial_\nu
A_\mu{}^a_m + f^{abc} A_\mu{}^b_n A_\nu{}^c_{m-n}                       
\end{eqnarray}
\begin{eqnarray}
\delta A_\mu{}^a_m &=& \partial_\mu \varepsilon^a_m + f^{abc}
A_\mu{}^b_n \varepsilon^c_{m-n}                   \nonumber   \\
\delta F_{\mu\nu}{}^a_m &=& f^{abc} F_{\mu\nu}{}^b_n
\varepsilon^c_{m-n}    \label{3}
\end{eqnarray}
\begin{equation}
{\cal L} = - \frac14 F_{\mu\nu}{}^a_m F_{\mu\nu}{}^a_{-m}.
\end{equation}

In order to consider spontaneous symmetry breaking, let us introduce
scalar field $\phi$, lying in algebra (\ref{1}):
\begin{equation}
\phi = \phi^a_m T^a_{-m}  \qquad  \phi^+ = - \phi  \qquad
(\phi^a_m)^* = \phi^a_{-m}
\end{equation}
Under the infinitesimal gauge transformations this field transforms
according to the usual rule:
\begin{equation}
\delta \phi = [\phi, \varepsilon]
\end{equation}
and covariant derivative has the form:
\begin{equation}
D_\mu \phi = [\nabla_\mu, \phi] = \partial_\mu \phi + [{\cal
A},\phi].
\end{equation}
In the components all these formulas take the following form:
\begin{eqnarray}
\delta\phi^a_m &=& f^{abc} \phi^b_n \varepsilon^c_{m-n} \label{5}  \\
D_\mu\phi^a_m &=& \partial_\mu \phi^a_m + f^{abc} A_\mu{}^b_n
\phi^c_{m-n}.
\end{eqnarray}

The total Lagrangian, invariant under the gauge transformations
(\ref{3}, \ref{5}), is the following:
\begin{equation}
{\cal L} = - \frac14 F_{\mu\nu}{}^a_m F_{\mu\nu}{}^a_{-m} + \frac12
D_\mu \phi^a_m D_\mu \phi^a_{-m}.                     \label{10}
\end{equation}

Now let us modify the gauge transformation of the field $\phi$.
Namely, let us introduce inhomogeneous term in transformations
(\ref{5}):
\begin{equation}
\delta\phi^a_m = f^{abc} \phi^b_n \varepsilon^c_{m-n} + i \mu m
\varepsilon^a_m     \label{13}
\end{equation}
Covariant derivative also changes its form:
\begin{equation}
D_\mu \phi^a_m = \partial_\mu \phi^a_m + f^{abc} A_\mu{}^b_n
\phi^c_{m-n} - i \mu m A_\mu{}^a_m                    \label{15}
\end{equation}
In this, Lagrangian (\ref{10}) with covariant derivative
$D_\mu\phi$, defined in (\ref{15}), is invariant under the gauge
transformations (\ref{3}, \ref{13}). Let us stress, that the fact we
are working with the infinite-dimensional algebra is crucial for the
possibility to have such a gauge invariance with inhomogeneous terms.
As we have already mentioned for any finite-dimensional algebra the
introduction of these inhomogeneous terms either leads to the
non-linear $\sigma$-models or requires the presence of the Higgs
fields.

It can be easily seen, that the vector fields $A_\mu{}^a_m$ with
$m\ne 0$ acquire masses, due to the following mass term (arising as
usual from the covariant derivatives in the scalar field kinetic
terms):
\begin{equation}
{\cal L}_M = \frac{\mu^2 m^2} 2 A_\mu{}^a_m A_\mu{}^a_{-m} =
\frac{\mu^2 m^2}2 A_\mu{}^a_m (A_\mu{}^a_m)^*.
\end{equation}
So, we have spontaneous breaking of the total Kac-Moody group to its
subgroup $G$, in this, the vector fields acquire masses,
proportional to the level number $m$ and to the symmetry breaking
scale $\mu$.

But we are interested in the fields from the lowest level, which we
associate with the observable particles. At this level gauge group
$G$ remains unbroken and corresponding vector fields remain massless.
In order to have spontaneous symmetry breaking, under which some of
the vector fields from the lowest level acquire masses, we should
generalize algebra (\ref{1}).

Let us assume, that group $G$ has some subgroup $H$ with generators
$T^a$, $a=1,...,{\rm dim}H$, all the other generators of the group
$G$ we denote as $T^{a'}$, $a'={\rm dim}H+1,...,{\rm dim}G$. Let the
commutation relations of this algebra are such that it admits a
${\bf Z}_2$-grading, i.e.:
\begin{equation}
[T^a, T^b] = f^{abc} T^c                  \qquad
[T^{a'}, T^{b'}] = f^{a'b'c} T^c          \qquad
[T^a, T^{b'}] = f^{ab'c'} T^{c'}
\end{equation}
For any such algebra it is not difficult to construct an infinite
dimensional algebra which will be the generalization of simplest case
described above. Namely, all the Jacoby identities will hold if one
assigns the integer levels to the generators of subgroup $H$ ---
$T^a_m$ and half-integer ones to other generators $T^{a'}_{m+1/2}$.
Corresponding commutation relations have the following form:
\begin{equation}
[T^a_m, T^b_n] = f^{abc} T^c_{m+n}                  \quad
[T^a_n, T^{b'}_{m+1/2}] = f^{ab'c'} T^{c'}_{m+n+1/2} \quad
[T^{a'}_{m+1/2}, T^{b'}_{n+1/2}] = f^{a'b'c} T^c_{m+n+1} \label{20}
\end{equation}

For the gauge field 
\begin{equation}
{\cal A}_\mu = A_\mu{}^a_m T^a_{-m} + A_\mu{}^{a'}_{m+1/2}
T^{a'}_{-(m+1/2)}
\end{equation} 
lying in algebra (\ref{20}), expressions for field strength and
the gauge transformations in the components are the following:
\begin{eqnarray}
F_{\mu\nu}{}^a_m &=& \partial_\mu A_\nu{}^a_m - \partial_\nu
A_\mu{}^a_m + f^{abc} A_\mu{}^b_n A_\nu{}^c_{m-n} + f^{ab'c'}
A_\mu{}^{b'}_{n+1/2} A_\nu{}^{c'}_{m-(n+1/2)}         \nonumber   \\
F_{\mu\nu}{}^{a'}_{m+1/2} &=& \partial_\mu A_\nu{}^{a'}_{m+1/2} -
\partial_\nu A_\mu{}^{a'}_{m+1/2} + f^{a'b'c} (A_\mu{}^{b'}_{m+1/2-n}
A_\nu{}^c_n - [\mu \leftrightarrow \nu])                  \label{23}
\end{eqnarray} 
\begin{eqnarray}
\delta A_\mu{}^a_m &=& \partial_\mu \varepsilon^a_m + f^{abc}
A_\mu{}^b_n \varepsilon^c_{m-n} + f^{ab'c'} A_\mu{}^{b'}_{n+1/2}
\varepsilon^{c'}_{m-(n+1/2)}                         \nonumber   \\
\delta A_\mu{}^{a'}_{m+1/2} &=& \partial_\mu \varepsilon^{a'}_{m+1/2}
+ f^{a'b'c} (A_\mu{}^{b'}_{n+1/2} \varepsilon^c_{m-n} - A_\mu{}^c_n
\varepsilon^{b'}_{m+1/2-n}).               \label{25}
\end{eqnarray}

Expressions for the covariant derivative and the gauge
transformations of the scalar field $\phi$, lying in algebra
(\ref{20}), in the components have the following form:
\begin{eqnarray}
D_\mu \phi^a_m &=& \partial_\mu \phi^a_m + f^{abc} A_\mu{}^b_n
\phi^c_{m-n} + f^{ab'c'} A_\mu{}^{b'}_{n+1/2} \phi^{c'}_{m-(n+1/2)} +
i \mu m \varepsilon^a_m              \label{28}   \\
D_\mu \phi^{a'}_{m+1/2} &=& \partial_\mu \phi^{a'}_{m+1/2} +
f^{a'b'c} (A_\mu{}^{b'}_{n+1/2} \phi^c_{m-n} - A_\mu{}^{c}_{n}
\phi^{b'}_{m+1/2-n}) + i \mu (m+1/2) \varepsilon^{a'}_{m+1/2}
\nonumber
\end{eqnarray}
\begin{eqnarray}
\delta \phi^a_m &=& f^{abc} \phi^b_n \varepsilon^c_{m-n} + f^{ab'c'}
\phi^{b'}_{n+1/2} \varepsilon^{c'}_{m_-(n+1/2)}       \nonumber   \\
\delta \phi^{a'}_{m+1/2} &=& f^{a'b'c} (\phi^{b'}_{n+1/2}
\varepsilon^c_{m-n} - \phi^c_n \varepsilon^{b'}_{m+1/2-n}.
\label{30}
\end{eqnarray}

The total Lagrangian, invariant under the gauge transformations
(\ref{25}, \ref{30}), is the following:
\begin{eqnarray}
{\cal L} &=& - \frac14 F_{\mu\nu}{}^a_m F_{\mu\nu}{}^a_{-m}
- \frac14 F_{\mu\nu}{}^{a'}_{m+1/2} F_{\mu\nu}{}^{a'}_{-(m+1/2)} +
\nonumber  \\
&& + \frac12 D_\mu \phi^a_m D_\mu \phi^a_{-m} + \frac12 D_\mu
\phi^{a'}_{m+1/2} D_\mu \phi^{a'}_{-(m+1/2)}.       \label{35}
\end{eqnarray}
The mass terms for the vector fields take the form: 
\begin{equation}
{\cal L}_M = \frac{\mu^2 m^2}2 A_\mu{}^a_m A_\mu{}^a_{-m}
+ \frac{\mu^2}2 (m+1/2)^2 A_\mu{}^{a'}_{m+1/2} A_\mu{}^{a'}_{m+1/2}.
\end{equation}

It is seen that from the fields of the lowest level the fields
$A_\mu{}^a_0$, lying in the subgroup $H$, remain massless, while the
fields $A_\mu{}^{a'}_{1/2}$ acquire masses $\mu/2$. Hence, such a
theory could indeed mimic the spontaneous gauge symmetry breaking $G
\rightarrow H$, for example, $SU(5) \rightarrow SU(3)\otimes SU(2)
\otimes U(1)$. In this, one would still have usual relations for
three gauge coupling constants.

\section{$N=2$ supersymmetry model}

As a preliminary step to a $N=2$ supergravity model let us consider
$N=2$ supersymmetry model with the mechanism of the gauge symmetry
breaking, described in the previous section. Here we are not
interested in the problem of the supersymmetry breaking and are
investigating, in which way vector fields acquire masses in a
supersymmetric model with a gauge Kac-Moody group.

There are two ways to describe a massive vector $N=2$ supermultiplet
\cite{Fay79}. In the first case the scalar Goldstone boson belongs to
another vector multiplet. This case leads to the so called massive
vector multiplets with central charge. As we have already mentioned,
in the $N=2$ supergravity the central charge will necessarily be
gauged, the gauge field being graviphoton. As the graviphoton plays a
very special role in our mechanism of spontaneous supersymmetry
breaking, we will not consider such multiplets in this paper.

In the second case Goldstone boson belongs to a hypermultiplet and
the central charge does not arise. To describe the corresponding
model let us consider some number of the vector multiplets $(A_\mu^M,
\Theta_i^M, {\cal Z}^M= {\cal X}^M+\gamma_5{\cal Y}^M)$ and
hypermultiplets $(\Omega^{iM}, X^M, L^{aM})$, carrying the same index
$M$, where $i=1,2$ and $a=1,2,3$. The scalar fields $X^M$ and $\vec
L^M$ of the hypermultiplet transform as a singlet and a triplet under
the $SU(2)$ automorphism group of the superalgebra. Such a
description of the hypermultiplets would enable us to consider the
fields $X^M$ as Goldstone ones without breaking the $SU(2)$
invariance.

The $N=2$ supersymmetric Lagrangian, before switching on the gauge
interactions, have the following form:
\begin{eqnarray}
{\cal L} &=& - \frac14 (A_\mu)^2 + \frac{i}2 \bar\Theta_i
\hat\partial \Theta_i + \frac12 \partial_\mu \bar{\cal Z}
\partial_\mu {\cal Z} +    \nonumber   \\
 && + \frac{i}2 \bar\Omega^i \hat\partial \Omega^i + \frac12
(\partial_\mu X)^2 + \frac12 (\partial_\mu \vec L)^2.     \label{40}
\end{eqnarray}
Supertransformations of the fields from both multiplets, under which
Lagrangian (\ref{40}) is invariant, are the following:
\begin{eqnarray}
 && \delta A_\mu = i (\bar\Theta_i \gamma_\mu \eta_i)     \qquad
\delta \Theta_i = - \frac12 (\sigma A) \eta_i - i \varepsilon_{ij}
\hat\partial {\cal Z} \eta_j                    \nonumber    \\
 && \delta {\cal X} = \varepsilon^{ij} (\bar\Theta_i\eta_j) \qquad
\delta {\cal Y} = \varepsilon^{ij} (\bar\Theta_i\gamma_5\eta_j)
\end{eqnarray}
\begin{eqnarray}
 && \delta \Omega^i = - i (\hat\partial X \delta_i{}^j + \hat\partial
L_i{}^j) \eta_i \nonumber   \\
 && \delta X = (\bar\Omega^i\eta_i)            \qquad
\delta \vec L = (\bar\Omega^i \vec\tau_i{}^j \eta_j),
\end{eqnarray}
where $\vec\tau_i{}^j$ are Pauli matrices and the following notation
was introduced: $L_i{}^j=\vec L\vec\tau_i{}^j$.

In order to switch on the gauge interaction in this model, let us
assume that all the fields from the vector and the hypermultiplets
transform under adjoint representation of some group $G$ with the
structural constants $f^{MNK}$. The following substitutions in
Lagrangian (\ref{40})  make this Lagrangian gauge invariant:
\begin{equation}
\partial_\mu {\cal Z}^M \rightarrow \partial_\mu {\cal Z}^M + f^{MNK}
A_\mu^N {\cal Z}^K
\end{equation}
with the analogous expressions for the other fields derivatives. In
order to restore the supersymmetry invariance, one has to add the
following terms to the Lagrangian and the supertransformation laws:
\begin{eqnarray}
{\cal L}' &=& f^{MNK} \left\{ - \frac12 \varepsilon^{ij}
(\bar\Theta_i^M {\cal Z}^N \Theta_j^K) + (\Theta^M_i X^N \Omega^{jK})
+ (\bar\Theta^M_i L^N_j{}^i \Omega^{jK}) + \frac12 \varepsilon_{ij}
(\bar\Omega^{iM} {\cal Z}^N \Omega^{jK}) \right\} +   \nonumber   \\
 && + \frac18 (f^{MNK} \bar{\cal Z}^N {\cal Z}^K)^2 - \frac12
|f^{MNK} X^N {\cal Z}^K|^2 - \frac12 |f^{MNK} \vec L^N {\cal Z}^K|^2
- \frac12 (\Delta^{aM})^2
\end{eqnarray}
\begin{eqnarray}
\delta'\Theta_i^M &=& \frac12 f^{MNK} \bar{\cal Z}^N {\cal Z}^K
\eta_i - \Delta^M_i{}^j \eta_j                       \nonumber   \\
\delta'\Omega^{iM} &=& \varepsilon^{ij} f^{MNK} {\cal Z}^N
(X^K\delta_j{}^k + L^K_j{}^k)\eta_k,
\end{eqnarray}
where the following notation is used:
\begin{equation}
\Delta^{aM}=f^{MNK}(X^NL^{aK}-\frac12\varepsilon^{abc}L^{bN}L^{cK})
\end{equation}

To demonstrate in this model the mechanism of the gauge symmetry
breaking, described in the previous section, let us assume that all
the fields lie in the Kac-Moody algebra (\ref{1}) rather than a
finite Lie algebra and divide index $M$ into a pair of indices
$\{A,m\}$, where $A$ is an index of adjoint representation of the
finite group $G$ and $m$ is an infinite index of the Kac-Moody
algebra. In this, structural constants take the form:
\begin{equation}
f^{MNK} = f^{ABC} \delta(m+n+k)
\end{equation}
and the summing rule has, for example, the following form:
$\partial_\mu \bar{\cal Z}^M\partial_\mu{\cal Z}^M = \partial_\mu
\bar{\cal Z}^A_m \partial_\mu {\cal Z}^A_{-m}$. Under the gauge
transformations all the fields except the fields $X^A_m$ transform
according to formulas (\ref{3}, \ref{5}) of the previous section
and transformation laws of the fields $X^A_m$ have an inhomogeneous
term (the same as in (\ref{13})):
\begin{equation}
 \delta X^A_m = f^{ABC} X^B_n \varepsilon^C_{m-n} + i \mu m
\varepsilon^A_m.
\end{equation}
In this, the covariant derivatives of the fields $X^A_m$ are the
following:
\begin{equation}
 D_\mu X^A_m = \partial_\mu X^A_m + f^{ABC} A_\mu{}^B_n X^C_{m-n} - i
\mu m A_\mu{}^A_m.     \label{45}
\end{equation}

In order to restore supersymmetry invariance, broken by the
inhomogeneous term in (\ref{45}), one has to add to the Lagrangian
and the supertransformation laws the following terms:
\begin{eqnarray}
 {\cal L}'' &=& i \mu m (\bar\Theta_i{}^A_m \Omega^i{}^A_{-m}) -
\frac{\mu^2m^2}2 \bar{\cal Z}^A_m {\cal Z}^A_{-m} - \frac{\mu^2m^2}2
\vec L^A_m \vec L^A_{-m}  -          \nonumber   \\
 && - \frac{i}2 (m-n) f^{ABC} ({\cal Z}^A_M \bar{\cal Z}^B_n + \vec
L^A_m\vec L^B_n) x^C_{-m-n}.
\end{eqnarray}
In the full correspondence with the non-supersymmetric model of the
previous section we have in the model under consideration a
spontaneous breaking of the gauge symmetry. Due to the supersymmetry
of the model, all the fields, both bosonic and fermionic ones,
acquire equal masses. The only exception is the fields $X^A_m$,
which turn out to be Goldstone ones. The mass terms of the model
look like:
\begin{equation}
 {\cal L}_M = \frac12 \mu^2 m^2 A_\mu{}^A_m A_\mu{}^A_{-m} + i \mu
m (\bar\Theta_i{}^A_m \Omega^i{}^A_{-m}) - \frac{\mu^2m^2}2 \bar{\cal
Z}^A_m {\cal Z}^A_{-m} - \frac{\mu^2 m^2}2 \vec L^A_m \vec L^A_{-m}.
\end{equation}
The invariance of the model under the supertransformations is intact
and all the fields can be grouped into the massive $N=2$
supermultiplets.

Now, it is an easy task to generalize the model considered to the
case of the generalized Kac-Moody algebra (\ref{20}). In this, a part
of the vector fields of the lowest level acquire masses and the gauge
group $G$ are broken to its subgroup $H$. Both scalar and spinor
fields acquire the same masses as the vector fields because the
supersymmetry is unbroken.

\section{$N=2$ supergravity model}

In this section we investigate the supergravity generalization of the
supersymmetric model described in the previous section. We choose to
work with a model of the $N=2$ supergravity interacting with vector
multiplets with the scalar field geometry $SO(2,m)/SO(2)\otimes
SO(m)$ and with hypermultiplets with the scalar fields geometry
$SO(4,m)/SO(4)\otimes SO(m)$. As it has been shown in \cite{Zin90},
such a combination of scalar field geometries admits a spontaneous
supersymmetry breaking with two arbitrary scales and without a
cosmological term. (Note, that such geometries appear in a natural
way in the investigations of $N=2$ $D=4$ superstrings). Moreover, the
$\sigma$-model chosen for the hypermultiplets is an essential part
of the models constructed in \cite{Tso96,Tso96a} where the scalar
fields parameterize non-symmetric quaternionic manifolds and,
therefore, it allows interesting generalizations.

\subsection{Vector multiplets}

To describe the interaction of vector multiplets with N=2
supergravity, let us introduce the following fields: graviton $e_{\mu
r}$, gravitini $\Psi_{\mu i}$, $i=1,2$, Majorana spinors $\rho_i$,
scalar fields $\hat{\varphi}$, $\hat{\pi}$, and $(m+2)$ vector
multiplets $\{A_\mu^M, \Theta_i^M,{\cal Z}^M ={\cal X}^M + \gamma_5
{\cal Y}^M \}$, $M=1,2,...m+2$, $g^{MN}=(--,+...+)$. It is not
difficult to see that the set of spinor and scalar fields is
superfluous (which is necessary for symmetrical description of
graviphotons and matter vector fields). The following set of
constraints corresponds to the model with the geometry
$SO(2,m)/SO(2)\otimes SO(m)$: 
\begin{equation}
\bar{\cal Z} \cdot {\cal Z} = - 2 \qquad {\cal Z} \cdot {\cal Z}=0
\qquad {\cal Z} \cdot \Theta_i = \bar{\cal Z} \cdot \Theta_i = 0.
\label{constr1}
\end{equation}

    The number of physical degrees of freedom is correct only
when the theory is invariant under the local $O(2) \approx  U(1)$
transformations, the combination $(\bar{\cal{Z}} \partial_\mu {\cal
Z})$ playing the role of a gauge field. Covariant derivatives for
scalar fields ${\cal Z}$ and $\bar{\cal{Z}}$ look like
\begin{equation}
 D_{\mu} = \partial_{\mu} \pm \frac12 (\bar{\cal{Z}} \partial_{\mu}
{\cal Z}),
\end{equation}
where covariant derivative $D_{\mu}{\cal Z}$ has the sign "+" and
$D_{\mu}\bar{\cal Z}$ has the sign "-".

    In the given notations the Lagrangian of the interaction looks as
follows:
\begin{eqnarray}
 {\cal L}^F &=& \frac{i}2 \varepsilon^{\mu\nu\rho\sigma}
\bar{\Psi}_{\mu i} \gamma_5 \gamma_\nu D_\rho \Psi_{\sigma i} +
\frac{i}2 \bar{\rho }^i \hat{D} \rho_i + \frac{i}2 \bar{\Theta }^i
\hat{D} \Theta_i - \nonumber \\
 && + e^{\hat\varphi /\sqrt{2}} \left\{ \frac14 \varepsilon^{ij}
\bar{\Psi}_{\mu i} ({\cal{Z}} (A^{\mu\nu} - \gamma_5
\tilde{A}^{\mu\nu})) \Psi_{\nu j} + \frac14 \bar{\Theta }^i
\gamma^\mu (\sigma A) \Psi _{\mu i} + \right. \nonumber \\
 && + \left. \frac{i}{4\sqrt{2}} \bar{\rho}^i \gamma^\mu ({\cal Z}
(\sigma A)) \Psi_{\mu i} + \frac{\varepsilon^{ij}}8 \left[ 2\sqrt{2}
\bar{\rho }_i (\sigma A) \Theta_j + \bar{\Theta}_i{}^M ({\cal Z}
(\sigma A)) \Theta_j{}^M \right] \right\}     \nonumber   \\
 && - \frac12 \varepsilon^{ij} \bar{\Theta}_i{}^M \gamma^\mu
\gamma^\nu D_\nu {\cal Z}^M \Psi_{\mu j} - \frac12 \varepsilon^{ij}
\bar{\rho}_i \gamma^\mu \gamma^\nu (\partial_\nu  \hat{\varphi}  +
\gamma_5 e^{-\sqrt{2}\hat{\varphi}} \partial_\nu \hat{\pi})
 \Psi_{\mu j} \label{vec_lf}   \\
 {\cal L}^B &=& - \frac12 R - \frac14 e^{\sqrt{2}\hat{\varphi}}
\left[ A_{\mu\nu}{}^2 + 2 ({\cal Z} \cdot A_{\mu\nu}) (\bar{\cal Z}
\cdot A_{\mu\nu}) \right] - \frac{\hat{\pi}}{2\sqrt{2}} (A \cdot
\tilde{A}) +   \nonumber \\
 &+& \frac12 (\partial_\mu \hat{\varphi})^2 + \frac12
e^{-2\sqrt{2}\hat{\varphi}} (\partial_\mu \hat{\pi})^2 + \frac12
D_\mu {\cal Z}^A D_\mu \bar{\cal Z}^A.  \label{vec_lb}
\end{eqnarray}

Covariant derivatives of the spinor fields have the following form:
\begin{eqnarray}
 D_\mu \eta_i &=& D_\mu^G \eta_i - \frac14 (\bar{\cal Z} \partial_\mu
{\cal Z}) \eta_i + \frac1{2\sqrt{2}} e^{-\sqrt{2}\hat{\varphi}}
\gamma_5 \partial_\mu \hat{\pi} \eta_i     \nonumber   \\
 D_\mu \rho_i &=& D_\mu^G \rho_i + \frac14 (\bar{\cal Z} \partial_\mu
{\cal Z}) \rho_i + \frac3{2\sqrt{2}} e^{-\sqrt{2}\hat{\varphi}}
\gamma_5 \partial_\mu \hat{\pi} \rho_i       \label{L1}    \\
 D_\mu \Theta_i &=& D_\mu^G \Theta_i - \frac14 (\bar{\cal
Z}\partial_\mu {\cal Z}) \Theta_i - \frac1{2\sqrt{2}}
e^{-\sqrt{2}\hat{\varphi}} \gamma_5 \partial_\mu \hat{\pi} \Theta_i
\nonumber
\end{eqnarray}
and derivative of the field $\Psi_{\mu i}$ is the same as for
$\eta_i$.

Supertransformation laws look like:
\begin{eqnarray}
 \delta \Theta^M_i &=& - \frac12 e^{\hat{\varphi}/\sqrt{2}}
\sigma^{\mu\nu} \left\{ A^M + \frac12 \bar{\cal Z}^M ({\cal Z} A) +
\frac12 {\cal Z}^M (\bar{\cal Z} A) \right\}_{\mu\nu} \eta_i - i 
\varepsilon_{ij} \hat{D} {\cal Z}^M \eta_i        \nonumber   \\
 \delta \rho_i &=& - \frac1{2\sqrt{2}} e^{\hat{\varphi}/\sqrt{2}}
{\cal Z} (\sigma A) \eta_i - i \varepsilon_{ij} \gamma^\mu
(\partial_\mu \hat{\varphi} + \gamma_5 e^{-\sqrt{2}\hat{\varphi}}
\partial_\mu \hat{\pi}) \eta_i          \nonumber  \\
 \delta \Psi{\mu i} &=& 2 D_\mu \eta_i + \frac{i}4 \varepsilon_{ij}
e^{\hat{\varphi}/\sqrt{2}} \bar{\cal Z} (\sigma A) \eta_i    \qquad
\qquad \delta \hat{\pi} = e^{\sqrt{2}\hat{\varphi}} \varepsilon^{ij}
(\bar{\rho}_i \gamma_5 \eta_j)            \nonumber     \\
 \delta {\cal X}^A &=& \varepsilon^{ij} (\bar{\Theta}_i{}^A \eta_j)
\qquad \delta {\cal Y}^A = \varepsilon^{ij} (\bar{\Theta}_i{}^A
\gamma_5 \eta_j)   \qquad \delta \hat{\varphi} = \varepsilon^{ij}
(\bar{\rho}_i \eta_j)    \nonumber    \\
 \delta A_\mu^A &=& e^{-\hat{\varphi}/\sqrt{2}} \left\{
\varepsilon^{ij} (\bar{\Psi}_{\mu i} {\cal Z}^A \eta_j ) + i
(\bar{\Theta}_i^A \gamma_\mu \eta^i) - \frac{i}{\sqrt{2}}
(\bar{\rho}^i \gamma_\mu {\cal Z}^A \eta_i) \right\}.  \label{TR1} 
\end{eqnarray}

\subsection{Hypermultiplets}

Now, in order to generalize the supersymmetric model of the previous
section, we need a parameterization of the $SO(4,m)/SO(m)\otimes
SO(4)$ non-linear $\sigma$-model where four scalar fields of the
hypermultiplet are divided into the singlet $X$ and the triplet $\vec
L$. Such a model has been constructed by the authors in \cite{Tso96}.
It contains, apart from the fields of $N=2$ supergravity, the
following fields: scalar field $\varphi$, Majorana spinor fields
$\chi^i$ and $(m+6)$ hypermultiplets $(X^A, \vec L^A, \Omega^{iA})$,
$g^{AB}=(-,-,-,+,...+)$, with the following constraints on the fields
$\vec L$ and $\Omega^i$, corresponding to the scalar field
geometry $SO(3,m+3)/SO(3)\otimes SO(m+3)$:
\begin{equation}
 L^{aA} L^b_A = - \delta^{ab}   \qquad  L^{aA} \Omega^i_A = 0
\label{constr2}
\end{equation}
This model is invariant under the local $SO(3)$-transformations with
the combination $A_\mu^a = \varepsilon^{abc} (L^{bA}
\stackrel{\leftrightarrow}{\partial} _\mu L^{cA})$, playing the role
of the gauge field. The corresponding covariant derivatives for the
fields $\vec L^A$, for example, have the following form:
\begin{equation}
 D_\mu L^{aA} = \partial_\mu L^{aA} + L^{bA} (L^{bB} \partial_\mu
L^{aB}) \qquad L^{aA} D_\mu L^b_A = 0
\end{equation}
As it has been shown in \cite{Tso96}, the scalar fields $(\varphi,
X^A, \vec L^A)$ parameterize quaternionic manifold with geometry
$SO(4,m+4)/SO(4) \otimes SO(m+4)$.

The Lagrangian of the model without the terms, describing the pure
$N=2$ supergravity, has the form:
\begin{eqnarray}
 {\cal L}_B &=& \frac12 (\partial_\mu \varphi)^2 + \frac12
e^{2\varphi} ((\partial_\mu X)^2 + 2 (\vec{L} \partial_\mu X)^2) +
\frac12 D_\mu \vec{L} D_\mu \vec{L} +                \nonumber  \\
 && + \frac{i}2 \bar\chi^i \hat D \chi^i + \frac{i}2 \bar\Omega^i
\hat D\Omega^i - \frac12 \bar{\Omega}^i \gamma^\mu \gamma^\nu
[ e^{\varphi} (\partial_\nu X + \vec{L} (\vec{L} \partial_\nu X))
\delta_i{}^j + D_\nu L_i{}^j ] \Psi_{\mu j} \nonumber \\
 && - \frac12 \bar{\chi}^i \gamma^\mu \gamma^\nu ( \partial_\nu
\varphi \delta_i{}^j - e^{\varphi} (L_i{}^j \partial_\nu X))
\Psi_{\mu j} + \frac{i}4 e^{\varphi} \varepsilon^{\mu\nu\rho\sigma}
\bar{\Psi}_\mu{}^i \gamma_5 \gamma_\nu (L_i{}^j \partial_\rho X)
 \Psi_{\sigma j} \nonumber \\
 && + \frac{i}4 e^{\varphi} \bar{\chi}_i \gamma^\mu (L_i{}^j
\partial_\mu X) \chi^j + \frac{i}4 e^{\varphi} \bar{\Omega}_i
\gamma^\mu (L_i{}^j \partial_\mu X) \Omega^j - i e^{\varphi}
\bar{\chi}_i \gamma^\mu \partial_\mu X \Omega^i \label{L2}
\end{eqnarray}
and the corresponding supertransformation laws are the following:
\begin{eqnarray}
 \delta \Psi_{\mu i} &=& 2 D_\mu \eta_i + e^{\varphi} (L_i{}^j
\partial_\mu X) \eta_j \nonumber \\
 \delta \chi^i &=& - i \gamma^\mu [ \partial_\mu \varphi \delta_i{}^j
- e^{\varphi} (L_i{}^j \partial_\mu X)] \eta_j \nonumber \\
 \delta \Omega^i &=& - i \gamma^\mu [ e^{\varphi} (\partial_\mu X +
\vec{L} (\vec{L} \partial_\mu X)) \delta_i{}^j + D_\mu L_i{}^j ]
\eta_j  \label{TR2}      \\
 \delta \varphi &=& (\bar{\chi}^i \eta_i) \quad \delta X =
e^{-\varphi} [(\bar{\Omega}^i \eta_i) + (\bar{\chi}^i L_i{}^j
\eta_j)] \quad \delta \vec{L} = (\bar{\Omega}^i (\vec{\tau})_i{}^j
\eta_j). \nonumber
\end{eqnarray}

If one adds an interaction with the vector multiplets, described in
the previous subsection, the following additional terms in the
Lagrangian arise:
\begin{eqnarray}
 \Delta{\cal L} &=& \frac{i}4 e^{\varphi} \{ (\bar\Theta_i \gamma^\mu
L_i{}^j \partial_\mu X \Theta_j) + (\bar\rho_i \gamma^\mu L_i{}^j
\partial_\mu X \rho_j) \} -   \nonumber  \\
 && - \frac{i}{4\sqrt 2} e^{\sqrt 2 \hat\varphi} \{ (\bar\chi^i
\gamma^\mu \gamma_5 \chi^i) + (\bar\Omega^i \gamma^\mu \gamma_5
\Omega^i) \} \partial_\mu\hat\pi - \nonumber   \\
 && - \frac18 e^{\hat\varphi/\sqrt 2} \{ (\bar\Omega^i \bar{\cal Z}
(\sigma A) \varepsilon_{ij} \Omega^j) + \bar\chi^i \bar{\cal Z}
(\sigma A)\varepsilon_{ij} \chi^j) \}.          \label{L3}     
\end{eqnarray}

In this, the whole Lagrangian (\ref{vec_lf}, \ref{vec_lb}, \ref{L2},
\ref{L3}) is invariant under the supertransformations (\ref{TR1},
\ref{TR2}).
 
\subsection{Spontaneous symmetry breaking}

The problem, we are interested in, is: if it is possible to break
simultaneously supersymmetry and gauge symmetry in the way, described
in the previous Section? Let us first consider the possibility of the
supersymmetry breaking. For this one has to detach the hidden sector
of the model and investigate its global symmetries. Let us divide the
index $M$ of the vector multiplets as $M=\{\tilde M, A\}$, $\tilde
M=1,2,3,4$, $g^{\tilde M \tilde N}=(-,-,+,+)$ and the index $\hat A$
of the hypermultiplets as $\hat A=\{\tilde A, A\}$, $\tilde
A=1,...,6$, $g^{\tilde A\tilde B}=(-,-,-,+,+,+)$. The hidden
sector contains the following fields from the vector multiplets:
$\rho_i$, $\hat\varphi$, $\hat\pi$ and $\{A_\mu^{\tilde M},
\Theta_i^{\tilde M}, {\cal Z}^{\tilde M}\}$ and the following fields
from the hypermultiplets: $\varphi$, $\chi^i$ and $\{X^{\tilde A},
\vec L^{\tilde A}, \Omega^{i\tilde A} \}$. The scalar fields from the
hypermultiplets, entering the hidden sector, parameterize the
quaternionic manifold $SO(4,4)/SO(4)\otimes SO(4)$, in this the
fields $X^{\tilde A}$ enter the Lagrangian through the divergency
only. In \cite{Zin90,Tso96a} it has been shown that the gauging of a
part of this global translations leads to the spontaneous
supersymmetry breaking with two arbitrary mass scales and vanishing
cosmological constant.

The observable sector of the model contains the vector multiplets
$(A_\mu, \Theta_i, {\cal Z})^A$ and the hypermultiplets $(X, \vec L,
\Omega^i)^A$. Let us assume, just like it have been made in the
previous section, that the fields from these multiplets lie in the
Kac-Moody algebra (\ref{1}) and divide index $A$: $A\to\{A,m\}$ with
$m$ being an infinite index. Then one can switch on the gauge
interaction in the observable sector with all the fields from both
types of the multiplets transforming under the same representation of
algebra (\ref{1}). For example, transformation laws for the
fields ${\cal Z}$ have the form:
\begin{equation}
 \delta {\cal Z}^A_m = f^{ABC} {\cal Z}^B_n \varepsilon^C_{m-n}
\end{equation}
and the same for all other fields, besides the fields $X^A_m$,
which have unhomogeneous term in the transformation laws:
\begin{equation}
\delta X^A_m = f^{ABC} x^B_n \varepsilon^C_{m-n} + i \mu m
\varepsilon^A_m
\end{equation}

The following substitutions into the Lagrangian of the model make it
gauge-invariant:
\begin{equation}
 \partial_\mu {\cal Z}^A_m \to \partial_\mu {\cal Z}^A_m + f^{ABC}
A_\mu{}^B_n {\cal Z}^C_{m-n}
\end{equation}
and similar ones for all the fields except $X^A_m$ and
\begin{equation}
 \partial_\mu X^A_m \to \partial_\mu X^A_m + f^{ABC} A_\mu{}^B_n
X^C_{m-n} - i \mu m A_\mu{}^A_m.
\end{equation}

As usual in supergravities, switching on the gauge interaction spoils
the invariance under the supertransformations and in order to restore
it one has to add the following terms to the Lagrangian and to the
supertransformation laws:
\begin{eqnarray}
 {\cal L}'^F &=& e^{-\hat\varphi/\sqrt 2} \left\{ - \frac14
\bar\Psi_{\mu i} \sigma^{\mu\nu} \varepsilon^{ij} \Delta_j{}^k
\Psi_{\nu k} + \frac{i}{2\sqrt 2} \bar\Psi_{\mu i} \gamma^\mu
\bar\Delta_i{}^j \rho_j - \frac{i}2 \bar\Psi_{\mu i} \gamma^\mu
\Delta^{\tilde M j}_i \Theta_j^{\tilde M} - \right. \nonumber \\
 && - \frac{i}2 \bar\Psi_{\mu i} \gamma^\mu (\Delta^A_m)_i{}^j
\Theta_j{}^A_{-m} - \frac{i}2 e^{\varphi} \bar\Psi_{\mu i} \gamma^\mu
\{ \bar\Delta^{\tilde M j}_i {\cal Z}^{\tilde M} +
(\bar\Delta_1{}^A_m)_i{}^j {\cal Z}^A_{-m}\} \varepsilon_{jk}\chi^k -
\nonumber \\
 && - \frac{i}2 e^{\varphi} \bar\Psi_{\mu i} \gamma^\mu
\varepsilon_{ij} \bar\Delta^{\tilde A} \Omega^{j\tilde A}
 + \frac{1}{\sqrt 2} \bar\rho_j \varepsilon^{jk} \Delta^{\tilde Mi}_k
\Theta^{\tilde M}_k - \nonumber \\
 && - \frac{i}2 \bar\Psi_{\mu i} \gamma^\mu \varepsilon_{ij} \{
e^{\varphi} f^{ABC} \bar{\cal Z}^B_n X^C_{m-n} \delta_k{}^j + f^{ABC}
\bar{\cal Z}^B_n (L^C_{m-n})_k{}^j + i e^{\varphi} \mu m 
\bar{\cal Z}^A_m\} \Omega^j{}^A_{-m} + \nonumber \\
 && + \frac1{\sqrt 2} \bar\rho_j \varepsilon^{jk} (\Delta^A_m)_k{}^i
\Theta_i{}^A_{-m} + \frac1{\sqrt2} \bar\rho_i \{ \bar\Delta^{\tilde
Mi}_j {\cal Z}^{\tilde M} + (\bar\Delta_1{}^A_m)_i{}^j {\cal
Z}^A_{-m} \} \chi^j - \frac2{\sqrt2} \bar\rho_i \bar\Delta^{\tilde A}
\Omega^{i\tilde A}    \nonumber \\
 && - \frac1{\sqrt2} \bar\rho_i \{ e^{\varphi} f^{ABC} \bar{\cal
Z}^B_n X^C_{m-n} \delta_j{}^i + f^{ABC} \bar{\cal Z}^B_n
(L^C_{m-n})_j{}^i - i e^{\varphi} \mu m \bar{\cal Z}^A_m \delta_j{}^i
\} \Omega^{jA}_{-m} - \nonumber \\
 && - \frac14 \bar\Theta^M_i \varepsilon^{i } \Delta_j{}^k \Theta^M_k 
- \bar\Theta_i^{\tilde M} \Delta^{\tilde Mi}_j \chi^j - e^{\varphi}
\bar\Theta^A_{-m} \{ (\Delta_1{}^A_m)_j{}^i + (\Delta_3{}^A_m)_j{}^i
\} \chi^j + \nonumber \\
 && + f^{ABC} \bar\Theta_i{}^A_{-m} \{ e^{\varphi} X^B_{m+n}
\delta_j{}^i + (L^B_{m+n})_j{}^i \} \Omega^{jC}_{-n}  - i e^{\varphi}
\mu m \bar\Theta_{im}^A \Omega^{iA}_{-m} - \frac14 \bar\chi^i \bar
{\tilde\Delta}_i{}^j \varepsilon_{jk} \chi^k - \nonumber \\
 && - \bar\chi^i \varepsilon_{ij} \bar\Delta^{\tilde A}
\Omega^{j\tilde A} - e^{\varphi} \bar\chi^i \varepsilon_{ij} \{
f^{ABC} \bar{\cal Z}^B_n X^C_{m-n} - i \mu m \bar{\cal Z}^A_m \}
\Omega^{jA}_{-m}  + e^{\varphi} \bar\Theta^{\tilde M}_i K^{\tilde
M\tilde A} \Omega^{i\tilde A} - \nonumber \\
 && - \frac14 \bar\Omega^{i\hat A} \bar{\tilde\Delta}_i{}^j
\varepsilon_{jk} \Omega^{k\hat A} + f^{ABC} \{ \frac12
\bar\Omega^i{}^A_m \varepsilon_{ij} \bar{\cal Z}^B_{-m-n}
\Omega^j{}^C_n  + \frac{i}4 \bar\Psi_{\mu i} \gamma^\mu {\cal Z}^B_n
\bar{\cal Z}^C_{-m-n} \Theta_i{}^A_m - \nonumber \\
 && \left. - \frac12 \bar\Theta_i{}^A_m \varepsilon^{ij} {\cal
Z}^B_{-m-n} \Theta_j{}^C_n + \frac1{2\sqrt2} \bar\Theta_i{}^A_m {\cal
Z}^B_{-m-n} \bar{\cal Z}^C_n \varepsilon^{ij} \rho_j \} \right\}
\end{eqnarray}
\begin{eqnarray}
{\cal L}'^B &=& - \frac12 e^{-\sqrt2 \hat\varphi} \left\{
\vec\Delta^{\tilde M}\vec \Delta^{\tilde M} + \vec\Delta^A_m
\vec\Delta^A_{-m} + 2 e^{2\varphi} | \vec\Delta^{\tilde M} 
{\cal Z}^{\tilde M} + \vec\Delta_1{}^A_m {\cal Z}^A_{-m} |^2 +
\right. \nonumber \\
 && + | \Delta^{ab}_2{}^A_m {\cal Z}^A_{-m} |^2 + e^{2\varphi} |
 {\cal Z}^{\tilde M} K^{\tilde M\tilde A} |^2  +  e^{2\varphi} |
f^{ABC} {\cal Z}^B_n X^C_{m-n} - i \mu  m {\cal Z}^A_m |^2 +
\nonumber \\
 && \left. + | f^{ABC} {\cal Z}^B_n \vec L^C_{m-n} |^2 + \frac14 |
f^{ABC} {\cal Z}^B_n \bar{\cal Z}^C_{m-n} |^2 \right\}
\end{eqnarray}
\begin{eqnarray}
\delta'\Psi_{\mu i} &=& e^{-\hat\varphi/\sqrt2} \frac{i}2 \gamma_\mu
\varepsilon^{ij} \Delta_j{}^k \eta_k         \qquad
\delta'\chi^i = - e^{\varphi} e^{-\hat\varphi/\sqrt2}
\varepsilon^{ij} \{ \Delta^{\tilde M}{}_k^j {\cal Z}^{\tilde M} +
(\Delta^A_m)_i{}^j {\cal Z}^A_{-m}\} \eta_k  \nonumber  \\
\delta'\rho_i &=& - e^{-\hat\varphi/\sqrt2} \frac1{\sqrt2}
\Delta_i{}^j \eta_j     \qquad
\delta'\Theta_i^{\tilde M} = e^{-\hat\varphi/\sqrt2} \{
\Delta^{\tilde M}{}_i^j + \frac12 {\cal Z}^{\tilde M}
\bar\Delta_i{}^j + \frac12 \bar{\cal Z}^{\tilde M} \Delta_i{}^j\}
\eta_j                  \nonumber   \\
\delta'\Omega^{i\tilde A} &=& \varepsilon^{ij}
e^{-\hat\varphi/\sqrt2} \{ {\cal Z}^{\tilde M} K^{\tilde M\tilde A}
\delta_i{}^j + L^{a\tilde A} [ \Delta^{a\tilde M} {\cal Z}^{\tilde M}
\delta_i{}^j + \Delta_1^a{}^A_n {\cal Z}^A_{-m} \delta_i{}^j -
\Delta_2^{ab}{}^A_m {\cal Z}^A_{-m} \tau^b{}_i{}^j ] \} \eta_j
\nonumber  \\
\delta'\Theta_i{}^A_m &=& e^{-\hat\varphi/\sqrt2} \{
(\Delta^A_m){}_i^j + \frac12 {\cal Z}^A_m \bar\Delta_i{}^j + \frac12
\bar{\cal Z}^A_m \Delta_i{}^j \} \eta_j + \frac12
e^{-\hat\varphi/\sqrt2} f^{ABC} {\cal Z}^B_n \bar{\cal Z}^C_{m-n}
\eta_i \nonumber   \\
\delta'\Omega^i{}_m^A &=& \varepsilon^{ij} e^{-\hat\varphi/\sqrt2} \{
e^{\varphi} [ f^{ABC} {\cal Z}^B_n X^C_{m-n} - i \mu m {\cal Z}^A_m +
\vec L^A_m (\vec\Delta^{\tilde M} {\cal Z}^{\tilde M} +
\vec\Delta_1{}^B_n {\cal Z}^B_{-n}) ] \delta_j{}^k + \nonumber  \\
 + && [ f^{ABC} {\cal Z}^B_n (L^C_{m-n})_i{}^j - L^a{}_m^A
\Delta_2^{ab} \tau^b{}_j{}^k ] \} \eta_k,
\end{eqnarray}
where the following notations are used:
\begin{eqnarray}
 \Delta^a &=& e^{\varphi} \Delta^{a\tilde M} {\cal Z}^{\tilde M} +
e^{\varphi} (\Delta_1^a)_m^A {\cal Z}^A_{-m} - \frac12
\varepsilon^{abc} (\Delta_2^{bc})^A_m {\cal Z}^A_{-m}  \nonumber   \\
 \tilde\Delta^a &=& e^{\varphi} \Delta^{a\tilde M} {\cal Z}^{\tilde
M} + e^{\varphi} (\Delta_1^a)_m^A {\cal Z}^A_{-m} + \frac12
\varepsilon^{abc} (\Delta_2^{bc})^A_m {\cal Z}^A_{-m}  \nonumber   \\
 \vec\Delta^{\tilde M} &=& K^{\tilde M\tilde A} \vec L^{\tilde A}
\qquad (\vec\Delta_1)^A_m = f^{ABC} X^B_n \vec L^C_{m-n} + i \mu m
\vec L^A_m      \nonumber  \\
 (\Delta_2^{ab})^A_m &=& f^{ABC} L^a{}^B_n L^b{}^C_{m-n}.
\end{eqnarray}

It is seen that the scalar field potential of the model has the
minimum corresponding to the vanishing vacuum expectation values of
the scalar fields from the observable sector, in this its value is
the following:
\begin{equation}
 V_0 = \frac12 <\{ (K^{\tilde M\tilde A} \vec L^{\tilde A})^2 + |
{\cal Z}^{\tilde M} K^{\tilde M\tilde A} |^2 + 2 | 
 {\cal Z}^{\tilde M} K^{\tilde M\tilde A} \vec L^{\tilde A} |^2 \} >.
\end{equation}
One can choose the following vacuum expectation values for the fields
of the hidden sector, consistent with the constraints on the scalar
fields, $<{\cal Z}^{\tilde M}>=(1,i,0,0)$ and $<L^{a\tilde A}> =
\delta^{a\tilde A}$. Also, let us choose the parameters $K^{\tilde
M\tilde A}$ of the local translations in the form: $K^{\tilde M\tilde
A} = M_1 \delta^{1\tilde M} \delta^{1\tilde A} + M_2\delta^{2\tilde
M}\delta^{2\tilde A}$. Now it is easy to check that vacuum
expectation value of the scalar potential equals zero, which
corresponds to the vanishing cosmological constant. In this, the
gravitini mass matrix takes the form:
\begin{equation}
 M^{ik} = - \frac12 \varepsilon^{ij} < \Delta_j{}^k> = \frac12
\pmatrix{M_1 + M_2 & 0 \cr 0 & M_1-M_2}      \label{M_G}
\end{equation}
and we have spontaneous supersymmetry breaking with two arbitrary
mass scales and, in particular, with the possibility of the partial
super-Higgs effect $N=2\to N=1$.

From the bosonic Lagrangian one can obtain, taking into account
constraints (\ref{constr1}) and (\ref{constr2}) on the fields ${\cal
Z}^M$ and $\vec L^{\hat A}$, the following mass terms for the scalar
fields of the model:
\begin{eqnarray}
{\cal L}_M^s &=& - \frac12 \left. M_1^2 L^1{}^A_m L^1{}^A_{-m} +
M_2^2 L^2{}^A_m L^2{}^A_{-m} + M_1^2 {\cal X}^A_m {\cal X}^A_{-m} +
M_2^2 {\cal Y}^A_m {\cal Y}^A_{-m} + \right\} \nonumber    \\
 && \left. + (\mu m)^2 [ \vec L^A_m \vec L^A_{-m} + {\cal Z}^A_m
\bar{\cal Z}^A_{-m} ] + 4 i \mu m [ M_1 L^1{}^A_m {\cal X}^A_{-m} +
M_2 L^2{}^A_m {\cal Y}^A_{-m} ] \right\}.
\end{eqnarray}
Corresponding mass terms for the fermionic and vector fields of the
observable sector are the following:
\begin{equation}
 {\cal L}_M^f = \frac12 \bar\Theta_i{}^A_m M^{ij} \Theta_j{}^A_{-m} -
i \mu m \bar \Theta_i{}^A_m \Omega^i{}^A_{-m} + \frac12
\bar\Omega^i{}^A_m M_{ij} \Omega^j{}^A_{-m},
\end{equation}
where the mass matrices $M_{ij}=M^{ij}$ are the same as in
(\ref{M_G}) and ${\cal L}_M^v = \frac12 (\mu m)^2 A_{\mu\nu}{}^A_m
A_{\mu\nu}{}^A_{-m}$.

After the diagonalization the mass spectrum of the model is the
following (see Table).
\begin{table}
\caption{The mass spectrum in the observable sector}
\begin{center}
\begin{tabular}{|c|c|c|c|}
\hline
 & vector fields & spinor fields & scalar fields \\
\hline 
      &           &                            &                \\
$m>0$ & $ \mu m $ & $\frac{M_1+M_2}2 + \mu m $ & $ M_1 + \mu m$
\\[6pt]
      &   & $ |\frac{M_1+M_2}2 - \mu m| $ & $ | M_1 - \mu m |$
\\[6pt]
      &   & $ |\frac{M_1-M_2}2 + \mu m| $ & $ M_2 + \mu m $   \\[6pt]
      &   & $ |\frac{M_1-M_2}2 - \mu m| $ & $ | M_2 - \mu m| $
\\[6pt]
      &   &                            &  $ \mu m $         \\[6pt]
\hline
      &        &                        &           \\
$m=0$ &  $ 0 $ &$ \frac{M_1+M_2}2   $   &  $ M_1$   \\[6pt]
      &        & $ |\frac{M_1-M_2}2| $   &  $ M_2 $  \\[6pt]
      &        &                         &   $ 0 $   \\[6pt]
\hline
\end{tabular}
\end{center}
\end{table}
At the lowest level ($m=0$) the vector fields are massless and for
each one we have two massless scalars, two scalars with masses equal
to $M_1$ and two ones with $M_2$ as well as two spinors with
masses equal to $(M_1+M_2)/2$ and the same number of spinors with
$(M_1-M_2)/2$. In the case of partial super-Higgs effect $N=2\to N=1$
($M_1=M_2=M$), all these fields form massless vector $N=1$
supermultiplets and massive (with masses equal to $M$) chiral $N=1$
supermultiplets, as it should be.

At the level with the level number $m$ for each massive vector field
with mass $(\mu m)$ we have pairs of scalar fields with masses
equal to $(M_1-\mu m)$, $(M_1+\mu m)$, $(M_2-\mu m)$, $(M_2+\mu m)$
and $(\mu m)$ and the pairs of spinor fields with masses equal to
$\frac{M_1+M_2}2 +\mu$, $\frac{M_1+M_2}2-\mu$, $\frac{M_1-M_2}2+\mu$
and $\frac{M_1-M_2}2-\mu$. Again, in the case when $N=2$
supersymmetry breaks to $N=1$ ($M_1=M_2=M$), all these fields form
massive vector $N=1$ supermultiplets with masses equal to $(\mu m)$
and the same number of massive scalar $N=1$ multiplets with masses
$(M+\mu m)$ and $(M-\mu m)$.

It is not difficult to obtain analogous results for the case of the
generalized Kac-Moody algebra (\ref{20}). In this case at the lowest
level part of the vector fields acquire masses, while the part of the
vector fields, associated with the generators of $H$ subgroup,
remains massless exactly in the same way as it would be when the
gauge symmetry breaks $G\to H$.

There is only one mass scale in the model with such mechanism of the
gauge symmetry breaking. Therefore, if one investigates a unified
model with a gauge group, such as $SU(5)$, then the offered scheme
can be used only to break the unification gauge group to the gauge
group of the Standard Model, for example, $SU(5)\to SU(3)\otimes
SU(2)\otimes U(1)$. Then one should again use the Higgs mechanism. 

One useful observation can be made from the mass spectrum of the
model under consideration. In the case, when, for example,
$\mu\approx M_1$, there is a number of the "light" scalar fields in
the model with masses equal to $(M_1-\mu)$. In the models with a
finite dimensional gauge group all such fields acquire masses $M_1$
and $M_2$ \cite{Zin90,Tso96a} which are close to the mass scale of
the supersymmetry breaking $N=2\to N=1$ and there is no way to lower
its values, while in the model considered a kind of inverse Higgs
mechanism is operative and these "light" particles can play a role
in the low energy phenomenology, for example, as the Higgs fields in
the breaking $SU(2)\otimes U(1)\to U(1)_{em}$.

\section*{Conclusion}

So we have managed to construct a class of $N=2$ supergravity models,
allowing simultaneous spontaneous breaking of both the supersymmetry
and the gauge symmetry. The supersymmetry was broken with two
arbitrary mass scales and vanishing cosmological constant. For the
gauge symmetry breaking specific properties of the gauge theories
with the infinite dimensional Kac-Moody algebras were used and it was
shown that such a mechanism worked in the case of $N=2$ supergravity.
It seems to be natural to use this scheme for the breaking of an
unification gauge group, such as $SU(5)$. One of the interesting
results, obtained as a byproduct of the whole construction, is that
after the gauge symmetry breaking some scalar fields, which after the
breaking of the supersymmetry acquire masses, close to the scale of
$N=2$ supersymmetry breaking, can be made "light" and can be used for
the breaking of the electro-weak gauge group like Higgs fields. For
this inverse Higgs mechanism to be operative, the mass scale of the
$N=2 \to N=1$ supersymmetry breaking have to be close to the scale of
unified gauge symmetry breaking exactly as $N=1 \to N=0$
supersymmetry breaking scale is expected to be close to electro-weak
one. Note, at last, that the quaternionic non-linear $\sigma$-model
we have chosen for the hypermultiplets allows one to consider the
generalization of our present work to the case of quaternionic
models,
based on non-symmetric quaternionic spaces, constructed in
\cite{Tso96,Tso96a}.

\vspace{0.3in}
{\large \bf Acknowledgements}
\vspace{0.2in}

Work supported by the International Science Foundation and Russian
Government grant RMP300 and by the Russian Foundation for Fundamental
research grant 95-02-06312.

\end{document}